\documentclass[11pt]{article}
\usepackage[margin=1in]{geometry}
\usepackage{graphicx}
\usepackage{amsmath}
\usepackage[numbers]{natbib}

\usepackage{hyperref}

\usepackage{lineno}
\usepackage{xcolor}
\usepackage{comment}
\usepackage[font=small]{caption}

\title{Modulation of the tropical meridional circulation by the Madden-Julian Oscillation}

\author{
Wuqiushi Yao\thanks{Corresponding author: wuqiushi.yao@weizmann.ac.il},
Or Hadas,
Yohai Kaspi\thanks{Corresponding author: yohai.kaspi@weizmann.ac.il}
}

\date{
Department of Earth and Planetary Science,  
Weizmann Institute of Science, Rehovot, Israel
}

\begin{document}
\maketitle

\begin{abstract}

The Hadley circulation is Earth’s dominant tropical overturning circulation, regulating atmospheric energy transport, tropical rainfall, and subtropical aridity. Although its variability has been extensively studied on seasonal to decadal climate-change timescales, its subseasonal behaviour remains poorly understood. Here we show that the Madden--Julian Oscillation (MJO), the leading mode of tropical intraseasonal variability, systematically modulates the Hadley circulation and influences global hydroclimate variability. Using reanalysis data and dynamical diagnostics, we identify a robust hemispherically asymmetric Hadley circulation anomaly associated with active MJO events, with amplitudes comparable to the climatological intraseasonal variability of the Hadley circulation. Breaking down the dynamical components reveals that the response is primarily maintained by latent heating associated with moist convection, while the overturning strength is driven by interhemispheric moisture gradients. Lead--lag analyses further show that the Hadley circulation lags the MJO by 4--15 days, indicating that MJO convection may drive overturning adjustments on subseasonal timescales. This coupled MJO--Hadley circulation state reveals a previously overlooked pathway linking tropical intraseasonal oscillations, meridional circulation and the hydrological cycle.

\end{abstract}

%
%
\newpage

\section*{Introduction}

In the tropics, where nearly 43\% of the global population resides\cite{StateOfTheTropics2020}, climate is strongly regulated by the Hadley circulation, the atmosphere’s dominant meridional overturning cell. The Hadley circulation is characterized by deep-tropical ascent of warm, moist air and intense precipitation, poleward flow in the upper troposphere, subtropical subsidence that defines the major desert belts, and a near-surface return flow toward the equator. Through this overturning, it redistributes energy and moisture across the tropics and subtropics, enhancing equatorial rainfall while suppressing precipitation in the descending branches.\\

Previous studies have extensively examined Hadley-cell variability on seasonal, interannual and decadal timescales. The seasonal evolution of the Hadley circulation is characterized by the transition between the hemispherically symmetric equinoctial circulation and the asymmetric solstitial circulation, with maximum intensity typically occurring in the winter hemisphere near solstice \citep{dima2003seasonality,stachnik2011comparison,YaoLuLiu2026}. On interannual timescales, ENSO strongly modulates the Hadley circulation \citep{oort1996observed,quan2004change,caballero2009impact,galanti2022spatial}, while on longer timescales substantial attention has focused on tropical widening, circulation weakening \citep{ChemkePolvani2019,ChemkeYuval2023}. In contrast, comparatively little attention has been paid to variability of the Hadley circulation on subseasonal timescales \citep{SchwendikeBerryEtAl2021}, despite the strong influence of subseasonal variability on hydroclimate, temperature extremes, agriculture, and food security.\\

Because the Madden--Julian Oscillation (MJO) is the dominant mode of tropical intraseasonal variability \citep{MaddenJulian1971,takasuka2021diversity,li2025distinct}, characterized by large-scale deep convection and circulation anomalies that propagates eastward from the Indian Ocean into the Pacific on timescales of roughly 30--90 days, it provides a natural candidate for driving subseasonal variability in the Hadley circulation. Previous studies have shown that the MJO induces pronounced regional overturning and divergent-wind anomalies over the Indo--Pacific warm pool \citep{SchwendikeBerryEtAl2021}. However, it remains unclear whether these regional circulation anomalies project coherently onto the zonal-mean Hadley circulation, or instead remain primarily localized responses. This distinction is important because the Hadley circulation is formally defined as a zonal-mean overturning circulation, whereas MJO convection is highly localized in longitude.\\

A substantial projection onto the zonal-mean circulation is therefore physically plausible. Although the Hadley circulation is defined as a zonal mean, its ascending branch is concentrated over the Indo--Pacific warm pool from both Eulerian \citep{YaoLuLiu2026,Schwendike2014} and Lagrangian perspectives \citep{galanti2022spatial,raiter2020tropical,HoskinsYang2023,raiter2024linking}. Raiter et al. \citep{raiter2020tropical,raiter2024linking} argued that the tropical overturning circulation is fundamentally organized by ascent over the warm pool and compensating subsidence elsewhere. Because the warm pool is also the primary region of MJO convective heating, moisture variability and diabatic forcing \citep{MajdaStechmann2009,YangIngersoll2013,SobelMaloney2012,SobelMaloney2013,AdamesKim2016}, MJO convection is expected to project strongly onto the zonal-mean Hadley circulation. Here we quantify the magnitude and structure of the resulting Hadley circulation response, identify the physical processes that maintain it, and assess its hydroclimatic impacts.\\

\begin{figure}[!htb]
    \centering
    \includegraphics[width=0.9\textwidth]{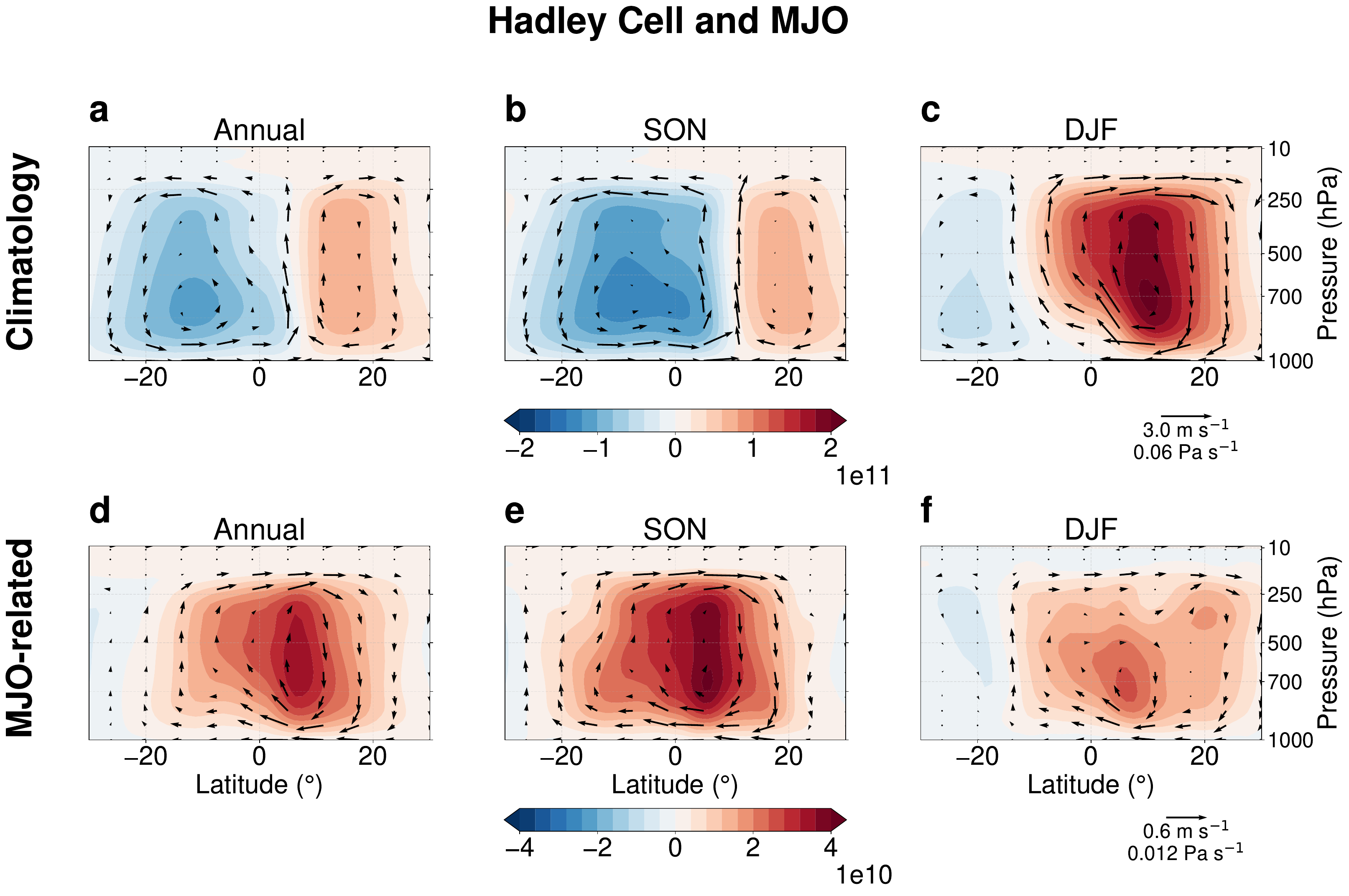}
    \caption{
    \textbf{Hadley circulation climatology, and its MJO-related anomaly.}
    (a--c) Climatological mean Hadley circulation for the annual mean, boreal autumn (SON), and boreal winter (DJF), respectively. The Hadley circulation is diagnosed using the meridional streamfunction (MSF; kg~s$^{-1}$; Methods). (d--f) Corresponding MJO-related Hadley circulation anomalies for the annual mean, SON, and DJF, defined as the difference between active and inactive MJO events (see Methods). Shading denotes the meridional streamfunction. Overlaid vectors indicate the predicted meridional and vertical circulation components $(v,\,-50\omega)$ derived from the streamfunction, where $v$ is the meridional wind (m~s$^{-1}$) and $\omega$ is the pressure vertical velocity (Pa~s$^{-1}$). The vertical velocity component is scaled by a factor of 50 for visualization.}
\end{figure}

\section*{MJO modulations on the Hadley circulation} 

To understand how the MJO modulates the Hadley circulation, we first examine the climatological Hadley circulation. The Hadley circulation is quantified using meridional streamfunction \cite{ChemkePolvani2019,raiter2020tropical,mitas2006recent} (MSF hereafter; Methods) The annual-mean Hadley circulation exhibits a dipole structure, with ascent in the deep tropics and subsidence in the subtropics of both hemispheres (Fig. 1a). The southern cell is slightly wider, consistent with the climatological position of the Intertropical Convergence Zone north of the equator \citep{SchneiderEtAl2014}. Boreal autumn resembles the annual-mean structure (Fig. 1b). In contrast, during boreal winter, the circulation becomes strongly asymmetric: the northern cell extends across the equator into the Southern Hemisphere, producing a much broader and stronger ascending branch compared to the weaker descending branch (Fig. 1c).\\

To quantify the influence of the MJO on this background state, we perform composite analyses of the MSF during active and inactive MJO periods (Methods). MJO-related Hadley circulation anomalies are diagnosed as the difference between active and inactive composites and are examined across seasons relative to their respective climatologies (Fig.~1 d,f). Across the annual mean, boreal autumn, and boreal winter, the MJO induces a coherent, hemispherically asymmetric circulation anomaly (Fig. 1d–f). This anomaly is characterized by anomalous ascent in the Southern subtropics, northward flow in the upper troposphere, subsidence in the Northern subtropics, and a compensating southward return flow near the surface, closely resembling the structure of the climatological boreal winter Hadley circulation (Fig.~1c). The MJO-related Hadley circulation anomaly reaches approximately 20\% of the climatological mean strength (Fig.~1) and nearly twice the magnitude of the intraseasonal standard deviation (\color{black}{Fig.~S2}\color{black}{}), indicating that a substantial fraction of tropical overturning variability on intraseasonal timescales is linked with the MJO rather than arising from stochastic variability. This result also suggests that the MJO constitutes a dominant and highly coherent mode of Hadley circulation variability. The robustness of this signal is confirmed across different declustering parameters (Fig.~S3; see also Methods), different MJO metrics (Fig.~S4), and multiple reanalysis datasets. In contrast, the MJO-related Hadley circulation anomaly during boreal spring and summer is substantially weaker and fails to form a coherent overturning structure (Fig.~S1). This weaker circulation response is consistent with the pronounced seasonal dependence of MJO activity itself, as the MJO convective signal is known to weaken and become less organized during boreal spring and summer (e.g., \cite{JiangEtAl2018}). We therefore focus primarily on boreal autumn and winter, when the MJO exerts its strongest and most dynamically coherent influence on the tropical overturning circulation.\\

\section*{A Kuo-Eliassen analysis of the modulated circulation}

The MJO-related Hadley circulation anomaly raises the question of what dynamical processes govern its formation. Two classes of mechanisms are plausible. One emphasizes momentum forcing, in which the circulation is driven by the convergence of eddy momentum fluxes associated with Rossby wave activity, as discussed in studies of tropical superrotation \citep{Arnold2012,Carlson2016}. The other highlights thermodynamic forcing, whereby hemispherically asymmetric latent heating associated with moist convection induces anomalous vertical motion and cross-equatorial mass transport \citep{AdamesMayta2024}.

To assess the relative importance of these processes, we employ the Kuo–Eliassen (KE) equation (Eq.~1, see also Methods):

\begin{equation}
\mathcal{L}(\Delta\psi)
=
\Delta S_{Q_{\mathrm{LH}}}+\Delta S_{Q_\mathrm{R}}
-
\Delta S_F
-
\Delta S_{\mathrm{ehf}}
-
\Delta S_{\mathrm{emf}}+\mathrm{RES},
\end{equation}
where $\mathcal{L}$ is a second-order elliptic operator acting on the anomalous MSF $\Delta \psi$.
The linearity of the KE equation allows the anomalous circulation response to be diagnostically decomposed into contributions from anomalous latent heating $\Delta S_{Q_{\mathrm{LH}}}$, radiative heating $\Delta S_{Q_\mathrm{R}}$, momentum damping $\Delta S_F$, eddy heat fluxes $\Delta S_{\mathrm{ehf}}$, eddy momentum fluxes $\Delta S_{\mathrm{emf}}$, and residual (RES) \citep{ChemkePolvani2019,ZaplotnikEtAl2022,YingEtAl2024,su2025consistency}. While the KE framework does not establish causality in non-steady flows \citep{HeldZuritaGotor2025}, it offers a physically consistent way to quantify the dominant dynamical balances. 

\begin{figure}[!htb]
    \centering
    \includegraphics[width=1.0\textwidth]{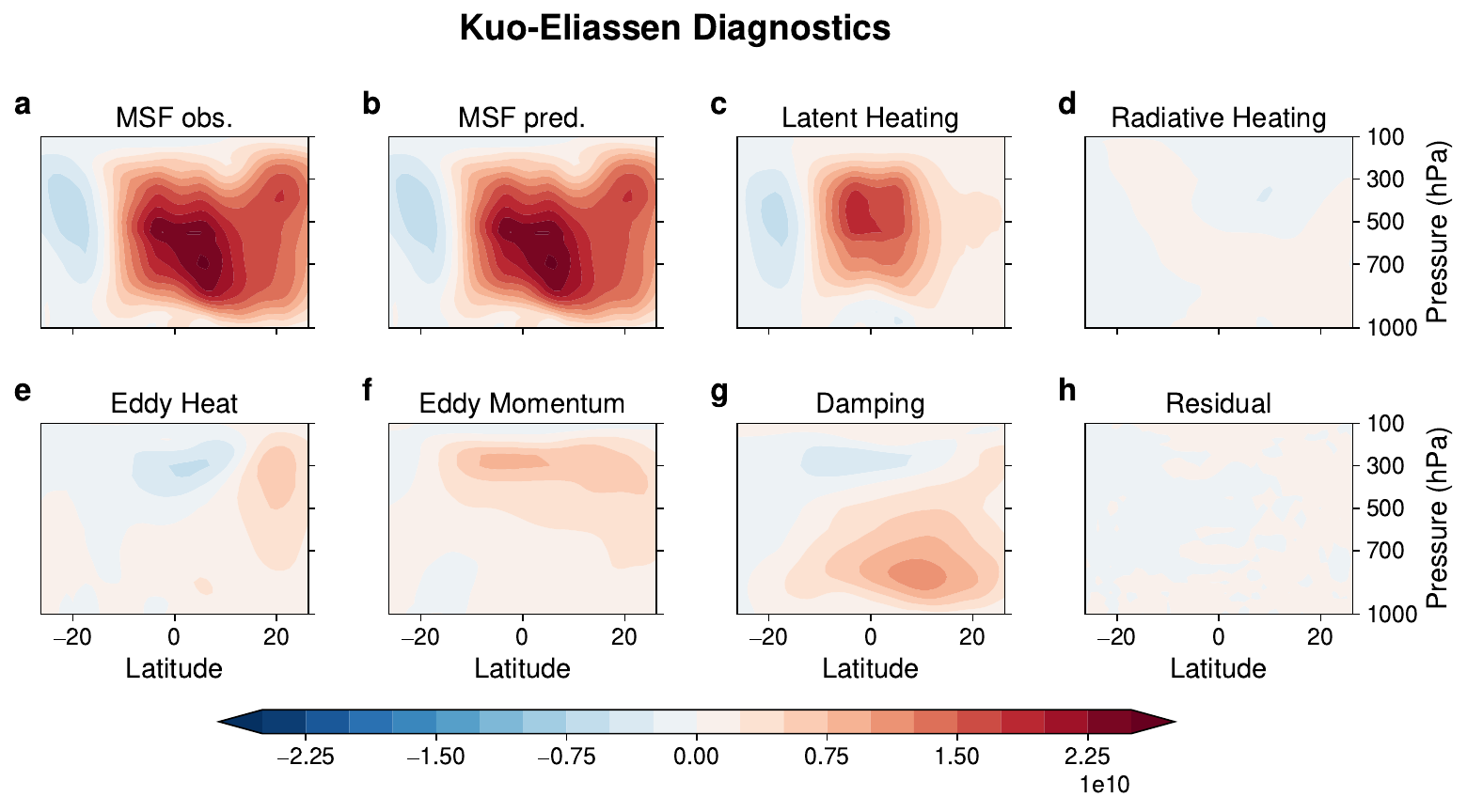}
    \caption{\textbf{Kuo–Eliassen (KE) equation diagnosis of MJO-related Hadley circulation variability in DJF.}
    (a–h) Differences between active and inactive MJO conditions in the MSF and the diagnosed KE components. Shown are the observed MSF difference (a), the KE-predicted MSF (b), and the individual forcing terms including latent heating (c), radiative heating (d), eddy heat flux convergence (e), eddy momentum flux convergence (f), momentum damping (g), and a residual term (h), defined as the difference between the total MSF change and the sum of the diagnosed components.
    Active (inactive) MJO conditions correspond to days in the upper (lower) 10\% of MJO amplitude. Pressure (hPa) is shown on the vertical axis and latitude (degrees) on the horizontal axis. Positive values denote an anomalous strengthening of the climatological Hadley circulation.
    }
    \label{fig:pdf_image}
\end{figure}

Figure~2 shows the KE diagnosis for boreal winter. The predicted MSF  well reproduces the observed anomaly (Fig.~2a,b), indicating that the KE equation well captures the observed circulation anomaly. The circulation anomaly is  dominated by latent heating (Fig.~2c), which accounts for the majority of the response. Eddy momentum flux convergence (Fig.~2f) and mechanical damping (Fig.~2g) provide secondary contributions, while eddy heat fluxes act to partially offset the total circulation (Fig.~2e). Similar results are obtained in boreal autumn (Fig.~S5), indicating that the dynamical balance is robust across seasons. In addition, convective processes can contribute to momentum damping through convective momentum transport \citep{LinEtAl2008}. Taken together, the KE diagnosis suggests that the Hadley circulation response is fundamentally constrained by moisture-related processes.

\section*{Scaling the intraseasonal Hadley circulation with the moisture gradient}

The KE diagnosis indicates that the MJO-related Hadley circulation anomaly is primarily controlled by the latent-heating term, which enters through the meridional gradient of diabatic heating. Because tropical diabatic heating is dominated by latent heat release associated with moist convection \citep{lachmy2020role}, with $Q \propto L_v q$, the anomalous Hadley circulation is therefore expected to scale with the meridional gradient of moisture. Consistent with this expectation, the zonally averaged latent-heating anomalies associated with strong MJO activity exhibit a pronounced hemispheric dipole structure in both SON and DJF (Fig.~3a,d), characterized by enhanced heating in the Southern Hemisphere and reduced heating in the Northern Hemisphere. A corresponding dipole structure is evident in the specific humidity anomalies (Fig.~3b,e), although the moisture anomalies peak at lower levels than the latent-heating anomalies. Together, these structures imply a substantial cross-equatorial gradient in both moisture and diabatic heating during active MJO conditions. Since the Hadley circulation is dynamically maintained by the meridional gradient of latent heating, this dipole structure suggests a direct link between the strength of the anomalous overturning circulation and the magnitude of the cross-equatorial moisture contrast.

To quantify this relationship, we define column-integrated moisture as
\begin{equation}
W=\frac{1}{g}\int q\,dp,
\end{equation}
and characterize the anomalous moisture gradient by the hemispheric contrast
\begin{equation}
\Delta W=W_{\mathrm{SH}}-W_{\mathrm{NH}},
\end{equation}
where $W_{\mathrm{SH}}$ and $W_{\mathrm{NH}}$ denote the area-averaged column moisture within the southern and northern boxes shown in Fig.~3c,f. For each $\Delta W$ bin, we compute the corresponding maximum anomalous Hadley circulation strength $\psi_{\max}$ at 500~hPa. In both SON and DJF, $\psi_{\max}$ scales approximately linearly with $\Delta W$ (Fig.~3c,f), indicating that stronger cross-equatorial moisture contrasts are associated with stronger anomalous overturning circulations. This empirical relationship is qualitatively consistent with the two-layer moisture-gradient theory of \cite{AdamesMayta2024}, which predicts that meridional circulation anomalies scale with the meridional gradient of column-integrated moisture. Although originally developed for the climatological mean Hadley circulation, the theory captures the large-scale structure of the MJO-related circulation anomaly reasonably well (Fig.~S6), suggesting that moisture gradients provide a useful organizing framework for understanding subseasonal Hadley circulation variability.

\begin{figure}[!htb]
    \centering
    \includegraphics[width=1.0\textwidth]{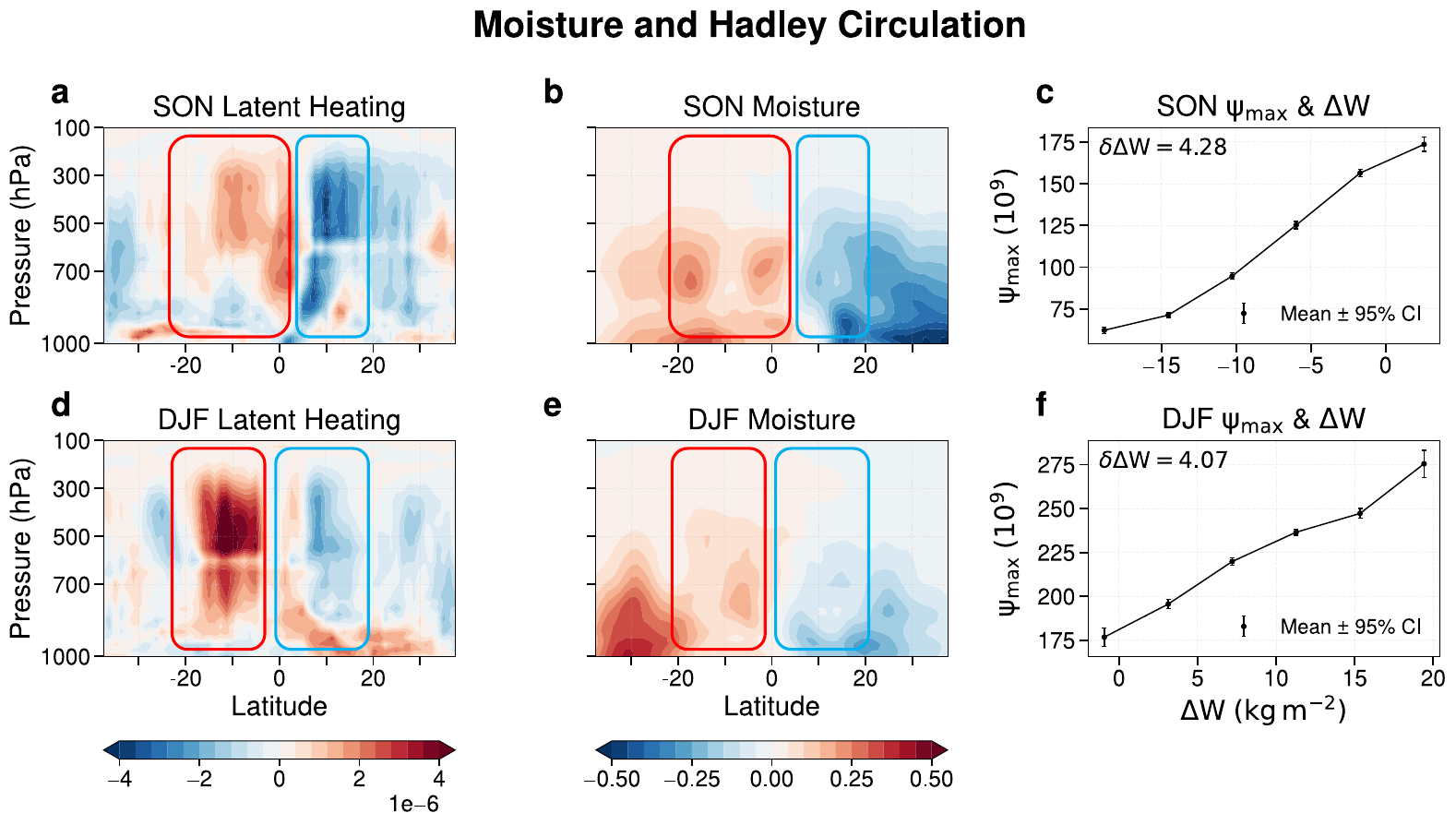}
    \caption{
    \textbf{Relationships between MJO-related thermodynamic anomalies and Hadley circulation variability.}
    (a,d) Zonally averaged latent heating anomalies ($Q$; K s$^{-1}$) in latitude–pressure coordinates for strong minus weak MJO events during SON and DJF, respectively. Positive values indicate enhanced diabatic heating during strong MJO activity.
    (b,e) Corresponding specific humidity anomalies ($q$; g kg$^{-1}$).
    (c,f) Binned relationship between the maximum Hadley circulation strength ($\psi_{\max}$; kg s$^{-1}$) and the hemispheric contrast in column-integrated moisture convergence ($\Delta W$; kg m$^{-2}$), defined as the moisture difference between the red box and blue box. Points denote binned means, solid lines connect adjacent bins, and error bars indicate 95\% confidence intervals (1.96 $\times$ standard error).
    }
    \label{fig:pdf_image}
\end{figure}

\section*{Lead-lag relationship between the MJO and the Hadley circulation}

\begin{figure}[!htb]
    \centering
    \includegraphics[width=0.8\textwidth]{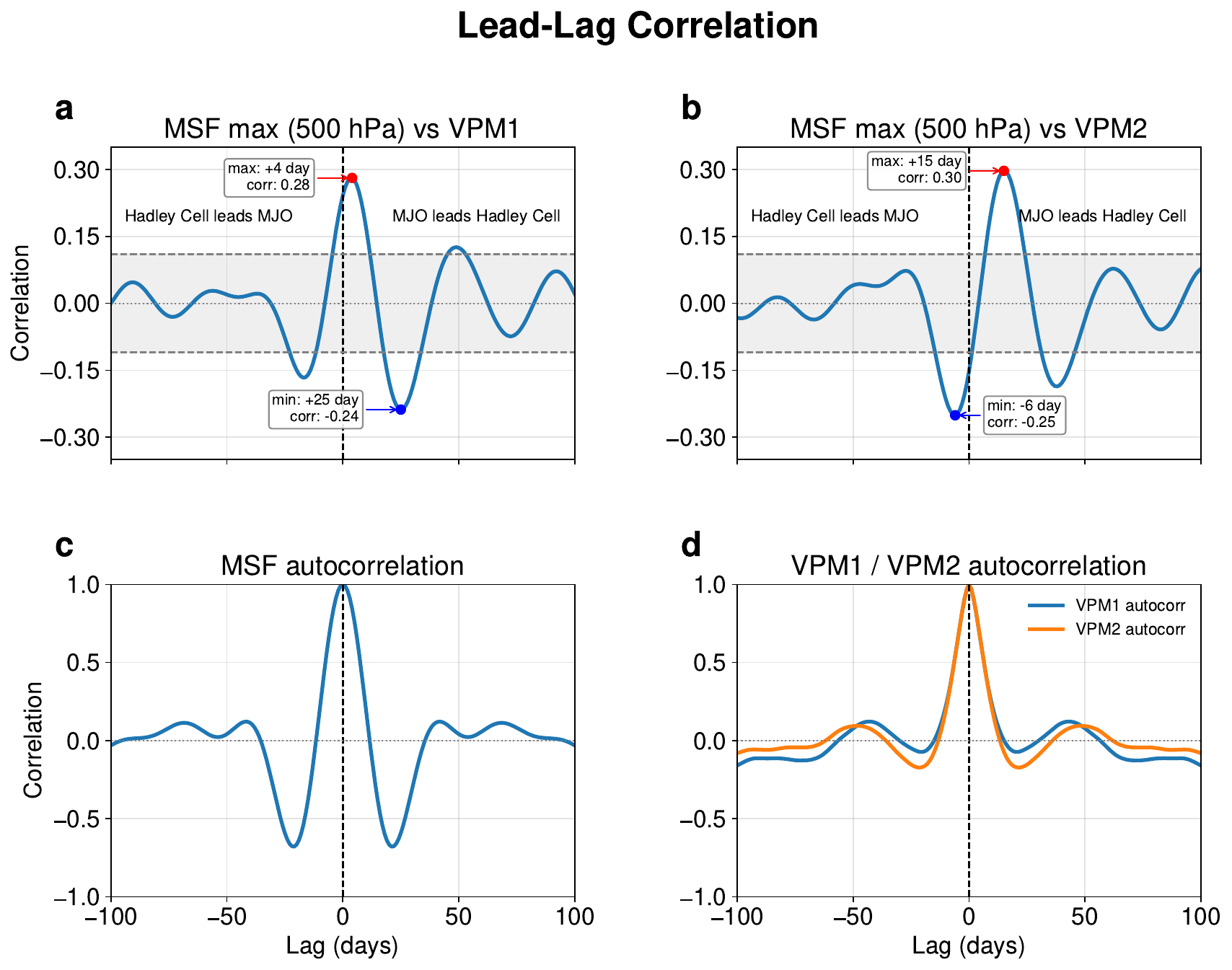}
    \caption{\textbf{Lead–lag relationships between the Hadley circulation (MSF) and the MJO index (VPM) at 500 hPa.}
    Lead--lag correlations between the maximum MSF between 25$\mathrm{^\circ S}$ and 25$\mathrm{^\circ N}$ at 500 hPa and the MJO indices are shown for (a) VPM1 and (b) VPM2. Positive lags indicate that the MJO leads the Hadley circulation. The gray shaded region denotes correlations within the 95\% confidence interval (\(|r| < 0.11\)), estimated using the effective sample size accounting for temporal autocorrelation (see Methods). Panels (c) and (d) show the autocorrelations of the 500-hPa MSF and the VPM indices, respectively. The MSF index is defined as the zonal-mean maximum MSF at 500 hPa after removal of the daily climatology and application of a band-pass filter. Red and blue markers in panel (a) denote the maximum positive and negative correlations, respectively. MSF is detrended and deseasoned, then 30-90 day bandpass filtered before computing the correlations.
}
    \label{fig:pdf_image}
\end{figure}
The thermodynamically driven nature of the MJO-related Hadley circulation anomaly raises a further question regarding causality: does the Hadley circulation actively modulate the MJO, or does the MJO force the intraseasonal Hadley circulation response? To address this issue, we examine lead–lag correlations between the maximum intraseasonal MSF at 500 hPa and the two-component MJO indices (Methods). This analysis provides a quantitative measure of the temporal relationship between the large-scale overturning circulation and MJO variability.

Figure~4a shows the lead–lag correlation between the Hadley-cell MSF index and VPM1. The correlation reaches a maximum of approximately 0.28 when the Hadley circulation lags VPM1 by about 4 days, whereas the strongest negative correlation occurs near +25 days with a magnitude of approximately $-0.24$. In contrast, the correlation maxima (Figure~4a: lag$\approx$-20) where the Hadley circulation leads the MJO, are substantially weaker. Similarly, the correlation between the Hadley-cell index and VPM2 peaks at approximately 0.30 when the Hadley circulation lags by about 15 days, while the minimum correlation occurs near $-6$ days with a magnitude of approximately $-0.25$ (Fig.~4b). The distinct lag between the VPM1 and VPM2 relationships is consistent with the well-known phase propagation of the MJO, in which the two indices represent different stages of eastward-propagating tropical convection.

The robustness of these relationships is further supported by the autocorrelation structure of the Hadley-cell index. Although the MSF autocorrelation exhibits strong persistence (Fig.~4c, compared with Fig.~4d, where the MJO index persistence is weak), with minima near $\pm 23$ days reaching correlations of approximately $-0.75$, the cross-correlations with the MJO indices are strongly asymmetric between positive and negative lags. This asymmetry indicates that the correlations cannot be explained solely by the intrinsic memory of the Hadley circulation itself. Instead, the systematic preference for positive lags demonstrates that the Hadley circulation responds to preceding MJO variability.

\section*{Discussion}

Taken together, the lead--lag relationship, the dominant role of latent heating, and the observed scaling with cross-equatorial moisture gradients suggest that the MJO-related Hadley circulation anomaly is fundamentally a moisture-coupled dynamical phenomenon. This raises a broader question regarding the dynamical mode that organizes the hemispherically asymmetric moisture anomalies and cross-equatorial overturning associated with the MJO. One possibility is that the MJO contains a significant contribution from mixed Rossby--gravity (MRG) wave packets \citep{takasuka2019mrg,takasuka2021mjo}. However, existing MRG-based theories are largely dry \citep{yang2011mjo,solodoch2011excitation} and do not explicitly represent moisture--convection coupling, whereas moisture-mode theories successfully describe moist convective feedbacks \citep{SobelMaloney2012,SobelMaloney2013,AdamesKim2016} but generally lack a representation of hemispherically asymmetric dynamics. Bridging these two perspectives could therefore provide a promising pathway toward a unified theory of tropical intraseasonal variability and its associated overturning circulations. The strong connection between moisture gradients, latent heating, and cross-equatorial overturning diagnosed here provides a potential observational constraint for such theories.

The circulation anomalies diagnosed here are also associated with substantial hydroclimatic impacts. Figure~S7 shows that the coupled MJO--Hadley circulation state produces coherent precipitation anomalies across large portions of the tropics and subtropics. During SON (Figure~S7a), the coupled state is associated with widespread drying over North Africa, the Sahel, India, Central Asia, and East Asia, accompanied by enhanced precipitation over parts of southern Africa and South America. During DJF (Figure~S7b), coherent drying persists over North Africa and South Asia, while enhanced precipitation emerges over Central America and portions of the southern United States. Although these patterns are broadly consistent with the known influences of both the MJO and the Hadley circulation on regional rainfall, comparison with composites based solely on MJO activity or Hadley circulation variability suggests that the coupled response is not simply a linear superposition of the two. In particular, the DJF coupled state exhibits precipitation anomalies that are substantially stronger and more spatially coherent than those associated with either active MJO conditions or strong Hadley circulation anomalies alone. This behaviour suggests that interactions between tropical convective variability and large-scale overturning circulation anomalies may amplify hydroclimatic responses on subseasonal timescales. Whether such coupling provides an additional source of subseasonal predictability remains an open question worthy of future investigation.

This study identifies a previously overlooked mode of subseasonal Hadley circulation variability that is systematically linked to the Madden--Julian Oscillation. Active MJO events induce a hemispherically asymmetric, boreal-winter-like overturning anomaly that rivals the climatological intraseasonal variability of the Hadley circulation and lags MJO by 4--15 days. Dynamical diagnostics indicate that the response is maintained primarily by latent heating associated with moist convection, while both empirical scaling and moist dynamical theory suggest that cross-equatorial moisture gradients provide a useful organizing framework for understanding the resulting circulation anomalies. Together, these results reveal how tropical moist convection organizes variability of the global overturning circulation on subseasonal timescales and identify MJO--Hadley circulation coupling as a fundamental mode of subseasonal tropical climate variability.


\color{black}{}

\newpage

\section*{Methods}
\label{sec:methods}
\subsection*{Reanalysis}
We use the ERA5 reanalysis dataset (horizontal resolution $1^\circ$; \citealp{hersbach2020era5}) and the NCEP--NCAR reanalysis dataset (horizontal resolution $2.5^\circ$; \citealp{kalnay2018ncep}) at 00, 06, 12, and 18 UTC during 1981--2020 to investigate the relationship between the MJO and subseasonal variations of the Hadley circulation. The analyzed variables include horizontal wind components ($u$, $v$), vertical pressure velocity ($\omega$), and specific humidity on pressure levels from 1000 to 10~hPa. The 6-hourly data are further averaged to obtain daily means. Unless otherwise noted, results shown in this study are based on ERA5, while the NCEP--NCAR reanalysis is used for validation.

\subsection*{MJO and Hadley circulation diagnostics}

The MJO is characterized using the Velocity Potential MJO (VPM) index \citep{ventrice2013modified}, which follows the EOF methodology of the Real-time Multivariate MJO (RMM) index \citep{wheeler2004all} but replaces outgoing longwave radiation with 200~hPa velocity potential. Because upper-level velocity potential emphasizes the large-scale divergent circulation associated with tropical deep convection, the VPM index is particularly suited for circulation-focused diagnostics of tropical overturning variability \citep{ventrice2013modified,kiladis2014comparison}. The MJO amplitude is defined as
\begin{equation}
A_\mathrm{VPM}\equiv \sqrt{\mathrm{VPM}_1^2+\mathrm{VPM}_2^2}.
\end{equation}

Rather than using fixed thresholds \citep{lafleur2015some}, active and inactive MJO conditions are identified using percentile-based thresholds \citep{dasgupta2020exploring}. Days with $A>A_{90\%}$ are classified as active MJO conditions, while days with $A<A_{10\%}$ are classified as inactive conditions. Seasonal analyses use thresholds computed separately for each season in order to account for seasonal variations in intraseasonal variability.

The Hadley circulation is quantified using the mass streamfunction (MSF, \citep{vallis2017atmospheric}):
\begin{equation}
\Psi(\phi,p)
=
\frac{2\pi a \cos\phi}{g}
\int_{0}^{p}
\overline{v}(\phi,p)\,dp ,
\label{eq:msf}
\end{equation}
where $\phi$ and $p$ denote latitude and pressure, respectively; $a$ is the Earth's radius; $g$ is the gravitational acceleration; and $\overline{v}$ is the zonal-mean meridional wind. The Hadley circulation strength is defined as the maximum value of the MSF at 500~hPa within $25^\circ\mathrm{S}$--$25^\circ\mathrm{N}$, and is used to quantify the relationship between tropical overturning variability and the MJO indices.

Daily MSF fields are composited separately during active and inactive MJO conditions, and the difference between these composites is defined as the MJO-related Hadley circulation anomaly. To avoid over-representing temporally persistent events, a declustering procedure is applied to the MJO amplitude time series prior to compositing. Strong (weak) events are identified from exceedances above the 90th percentile (below the 10th percentile), and consecutive exceedance days are grouped into individual event clusters.

To assess the sensitivity of the diagnosed anomalies to event persistence, we further introduce a persistence-weighting parameter, $\alpha$, which controls the weighting of consecutive exceedance days within each cluster (Fig.~S3). Here, $\alpha = 1$ treats every exceedance day as an independent event, whereas $\alpha = 0$ assigns equal weight to each event cluster regardless of duration. Figure~S3 shows that varying $\alpha$ changes the magnitude of the diagnosed MSF anomalies quantitatively but does not alter their overall spatial structure, indicating that the main results are robust to the treatment of event persistence.

\subsection*{Kuo--Eliassen equation}

To diagnose the physical processes responsible for Hadley circulation variability, we employ the Kuo--Eliassen (KE) equation, a linear second-order elliptic partial differential equation for the meridional streamfunction (MSF) derived from the quasi-geostrophic momentum and thermodynamic equations \citep{krishnamurti1979tropical}:

\begin{equation}
\mathcal{L}\psi
=
S_Q
-
S_F
-
S_{\mathrm{ehf}}
-
S_{\mathrm{emf}},
\end{equation}

\noindent where $\mathcal{L}$ is the KE operator,

\begin{equation}
\mathcal{L}
=
f^{2}\,\frac{g}{2\pi a \cos\phi}\,\frac{\partial^{2}}{\partial p^{2}}
+
\Gamma\,\frac{g}{2\pi a}\,
\frac{\partial}{a\,\partial\phi}
\left(
\frac{1}{a\cos\phi}
\frac{\partial}{\partial\phi}
\right).
\end{equation}

\noindent Here $f$ is the Coriolis parameter, $g$ is gravitational acceleration, $a$ is Earth's radius, $\phi$ is latitude, and $p$ is pressure. The static stability is defined as
$\Gamma=-\frac{R_d}{p}\left(\frac{\partial[T]}{\partial p}-\frac{R_d}{c_p}\frac{[T]}{p}\right)$,
where $T$ is temperature, $R_d$ is the gas constant for dry air, and $c_p$ is the specific heat capacity at constant pressure.

The forcing terms represent diabatic heating ($S_Q$), friction ($S_F$), eddy heat fluxes ($S_{\mathrm{ehf}}$), and eddy momentum fluxes ($S_{\mathrm{emf}}$), respectively. The diabatic-heating forcing is
$S_Q=\frac{R_d}{p}\frac{\partial [Q]}{a\partial\phi}$,
where $Q$ is the total diabatic heating. We further decompose
$Q=Q_{\mathrm{LH}}+Q_{\mathrm{R}}$
into latent-heating and radiative-heating components, such that
$S_Q=S_{Q_{\mathrm{LH}}}+S_{Q_{\mathrm{R}}}$.
The frictional forcing is
$S_F=f\frac{\partial[D]}{\partial p}$,
where $D$ represents zonal friction. The eddy heat-flux forcing is
$S_{\mathrm{ehf}}=\frac{R_d}{p}\frac{\partial}{a\partial\phi}\left[\frac{1}{a\cos\phi}\frac{\partial}{\partial\phi}([v'T']\cos\phi)\right]$,
where $v'T'$ is the eddy heat flux. The eddy momentum-flux forcing is
$S_{\mathrm{emf}}=
f\frac{1}{a\cos^{2}\phi}
\frac{\partial^{2}}{\partial p\partial\phi}
([u'v']\cos^{2}\phi)$,
where $u'v'$ is the eddy momentum flux.

In this study we apply the KE equation to composite anomalies associated with MJO activity:

\begin{equation}
\mathcal{L}(\Delta\psi)
=
\Delta S_{Q_{\mathrm{LH}}}
+
\Delta S_{Q_{\mathrm{R}}}
-
\Delta S_F
-
\Delta S_{\mathrm{ehf}}
-
\Delta S_{\mathrm{emf}}
+
\mathrm{RES},
\end{equation}

\noindent where $\Delta$ denotes the difference between composites of active and inactive MJO events. This decomposition exploits the linearity of the KE equation. Strictly speaking,
$\Delta(\mathcal{L}\psi)=\mathcal{L}(\Delta\psi)+\Delta\mathcal{L}(\psi)$,
where $\Delta\mathcal{L}$ arises primarily from changes in static stability between active and inactive MJO conditions. In our analysis this contribution is small and is therefore included within the residual term (RES), together with numerical errors and departures from the quasi-geostrophic approximation (Fig.~2h and Fig.~S5h).

\subsection*{Scaling the Hadley circulation with the moisture gradient}

From the KE diagnosis, the latent-heating term dominates the anomalous MSF. Assuming that latent heating is proportional to moisture, $Q_{\mathrm{LH}}\approx L_vq$, where $q$ is specific humidity and $L_v$ is the latent heat of vaporization, the KE equation implies $\mathcal{L}\psi\approx \partial[Q_{\mathrm{LH}}]/\partial y \approx L_v\partial q/\partial y$. If the vertical-curvature term dominates the KE operator, then $\mathcal{L}\psi\propto\partial^2\psi/\partial p^2$, yielding

\begin{equation}
[v]\propto L_v\frac{\partial W}{\partial y},
\end{equation}

\noindent where
$W=\frac{1}{g}\int q\,dp$
is the column-integrated moisture and we have used the relation $\partial\psi/\partial p\propto[v]$. This scaling suggests that stronger meridional moisture gradients should be associated with stronger anomalous overturning circulations.

This result is qualitatively consistent with the two-layer moist dynamical framework \cite{AdamesMayta2024}. In this model, the lower-tropospheric meridional wind varies parabolically across the Hadley-cell domain, while the upper-tropospheric wind has equal magnitude and opposite sign. The circulation amplitude is

\begin{equation}
[\overline{v}_0]
=
\frac{L_H^2(1+r)}{2S\tau_c}
\frac{\partial L_v[\overline{W}]}{\partial y},
\end{equation}

\noindent where $L_H$ is the Hadley-cell half-width, $r$ is the cloud-radiative feedback parameter, $S$ is the gross dry stability, $\tau_c$ is the convective moisture adjustment timescale, and $[\overline W]$ is the zonal-mean column-integrated moisture. The theory therefore predicts the same leading-order dependence of circulation strength on the meridional moisture gradient as the KE-based scaling derived above.

We evaluate the theoretical circulation using both the canonical parameter choice $\tau_c=4$ days and a shorter adjustment timescale $\tau_c=2$ days, which is more consistent with observational estimates of tropical convective adjustment timescales (Fig.~S6, \cite{AdamesMayta2024}).

\subsection*{Lead--lag correlations and significance levels}

Lead--lag correlations are calculated between the MJO indices (VPM1 and VPM2) and the Hadley circulation strength, defined as the zonal-mean maximum MSF between $25^\circ\mathrm{S}$ and $25^\circ\mathrm{N}$ at 500~hPa (Fig.~4). The analysis uses the full 40-year daily time series. Prior to the calculation, the MSF time series are detrended, deseasonalized by removing the daily climatology, and band-pass filtered using a 30--90 day filter in order to isolate subseasonal variability associated with the MJO. The same detrending and deseasonalization procedures are applied to the VPM indices before computing the correlations. Lead--lag correlations are then calculated by stepwise shifting the two time series relative to each other over a lag range of $\pm 60$ days. Positive lags indicate that the MJO leads the Hadley circulation response.

The statistical significance of the correlations is assessed using a two-tailed Student's $t$-test. Because both the MSF and VPM indices exhibit strong temporal autocorrelation, the number of independent samples is smaller than the total number of daily samples. An effective sample size ($N_{\mathrm{eff}}$, \cite{bretherton1999effective}) is therefore estimated :
\begin{equation}
N_{\mathrm{eff}} = N \frac{1-r_1 r_2}{1+r_1 r_2},
\end{equation}
where $N$ is the total sample size, and $r_1$ and $r_2$ are the lag-1 autocorrelations of the two time series. Using lag-1 autocorrelations of $r_1 = 0.99$ for the MSF and $r_2 = 0.968$ for the VPM indices, the effective sample size was estimated to be approximately $N_{\mathrm{eff}} = 311$.

Based on the effective degrees of freedom, the corresponding two-tailed significance thresholds are approximately $|r| \approx 0.11$ for the 95\% confidence level and $|r| \approx 0.15$ for the 99\% confidence level. Correlations exceeding these thresholds are considered statistically significant.

\section*{Data availability}

The Real-time Multivariate MJO (RMM) index used in this study is publicly available from the Australian Bureau of Meteorology (BoM):
\url{https://www.bom.gov.au/climate/mjo/graphics/rmm.74toRealtime.txt}.\\

The Velocity Potential MJO (VPM) index is obtained from the NOAA Physical Sciences Laboratory (PSL) MJO archive:
\url{https://psl.noaa.gov/mjo/mjoindex/}.\\

The ERA5 reanalysis dataset \citep{hersbach2020era5}, spanning from 1979 to the present, is publicly available from the Copernicus Climate Data Store at:
\url{https://cds.climate.copernicus.eu/cdsapp#!/dataset/reanalysis-era5-pressure-levels?tab=form}.\\

The NCEP-NCAR reanalysis dataset \citep{kalnay2018ncep}
, spanning from 1979 to the present, is publicly available from the Copernicus Climate Data Store at:
\url{https://psl.noaa.gov/data/gridded/data.ncep.reanalysis.html}.

\newpage

\bibliography{agusample}

@article{AdamesMayta2024,
  author  = {Adames, {\'A}ngel F. and Mayta, V. C.},
  title   = {The stirring tropics: Theory of moisture mode--{Hadley} cell interactions},
  journal = {Journal of Climate},
  year    = {2024},
  volume  = {37},
  number  = {4},
  pages   = {1383--1401},
  doi     = {10.1175/JCLI-D-23-0192.1}
}

@article{ChemkeYuval2023,
  author  = {Chemke, R. and Yuval, J.},
  title   = {Human‐induced weakening of the {Northern Hemisphere} tropical circulation},
  journal = {Nature},
  year    = {2023},
  volume  = {617},
  number  = {7961},
  pages   = {529--532},
  doi     = {10.1038/s41586-023-05903-1}
}

@article{ChemkePolvani2019,
  author  = {Chemke, R. and Polvani, L. M.},
  title   = {Opposite tropical circulation trends in climate models and in reanalyses},
  journal = {Nature Geoscience},
  year    = {2019},
  volume  = {12},
  number  = {7},
  pages   = {528--532},
  doi     = {10.1038/s41561-019-0383-x}
}

@article{dima2003seasonality,
  title={On the seasonality of the Hadley cell},
  author={Dima, I. M. and Wallace, J. M.},
  journal={Journal of the Atmospheric Sciences},
  volume={60},
  number={12},
  pages={1522--1527},
  year={2003}
}

@article{stachnik2011comparison,
  title={A comparison of the {Hadley circulation} in modern reanalyses},
  author={Stachnik, J. P. and Schumacher, C.},
  journal={Journal of Geophysical Research: Atmospheres},
  volume={116},
  number={D22},
  year={2011},
  publisher={Wiley Online Library}
}

@article{oort1996observed,
  title={Observed interannual variability in the {Hadley} circulation and its connection to {ENSO}},
  author={Oort, A. H. and Yienger, J. J.},
  journal={Journal of Climate},
  pages={2751--2767},
  year={1996},
  publisher={JSTOR}
}

@article{JiangEtAl2018,
  author  = {Jiang, X. and Adames, {\'A}lvaro F. and Zhao, Ming and Waliser, Duane and Maloney, Eric},
  title   = {A unified moisture mode framework for seasonality of the {Madden--Julian Oscillation}},
  journal = {Journal of Climate},
  volume  = {31},
  number  = {11},
  pages   = {4215--4224},
  year    = {2018},
  doi     = {10.1175/JCLI-D-17-0671.1}
}

@incollection{quan2004change,
  title={Change in the tropical {Hadley cell} since 1950},
  author={Quan, Xiao-Wei and Diaz, Henry F and Hoerling, Martin P},
  booktitle={The Hadley Circulation: Present, Past and Future},
  pages={85--120},
  year={2004},
  publisher={Springer}
}

@article{HoskinsYang2023,
  author  = {Hoskins, B. J. and Yang, G-Y.},
  title   = {A Global Perspective on the Upper Branch of the Hadley Cell},
  journal = {Journal of Climate},
  year    = {2023},
  volume  = {36},
  number  = {19},
  pages   = {6749--6762},
  doi     = {10.1175/JCLI-D-22-0537.1}
}

@article{caballero2009impact,
  title={Impact of midlatitude stationary waves on regional {Hadley cells and ENSO}},
  author={Caballero, Rodrigo and Anderson, Bruce T},
  journal={Geophysical Research Letters},
  volume={36},
  number={17},
  year={2009},
  publisher={Wiley Online Library}
}

@article{MaddenJulian1971,
  author  = {Madden, Roland A. and Julian, Paul R.},
  title   = {Detection of a 40--50 day oscillation in the zonal wind in the tropical {Pacific}},
  journal = {Journal of the Atmospheric Sciences},
  year    = {1971},
  volume  = {28},
  number  = {5},
  pages   = {702--708},
  doi     = {10.1175/1520-0469(1971)028<0702:DOADOO>2.0.CO;2}
}

@article{Schwendike2014,
  author       = {Schwendike, Juliane and Govekar, Pallavi and Reeder, Michael J. and Wardle, Richard M. and Berry, Gareth J. and Jakob, Christian},
  title        = {Local partitioning of the overturning circulation in the tropics and the connection to the {Hadley and Walker} circulations},
  journal      = {Journal of Geophysical Research: Atmospheres},
  volume       = {119},
  number       = {3},
  pages        = {1322--1339},
  year         = {2014},
  doi          = {10.1002/2013JD020742},
}

@book{vallis2017atmospheric,
  title={Atmospheric and oceanic fluid dynamics},
  author={Vallis, Geoffrey K},
  year={2017},
  publisher={Cambridge University Press}
}

@report{StateOfTheTropics2020,
  title={State of the Tropics 2020 Report},
  author={Howden, S. M. and Crimp, S. J. and Stokes, C. J. and others},
  institution={James Cook University},
  address={Cairns, Australia},
  year={2020}
}

@article{MajdaStechmann2009,
  author  = {Majda, Andrew J. and Stechmann, Samuel N.},
  title   = {The skeleton of tropical intraseasonal oscillations},
  journal = {Proceedings of the National Academy of Sciences},
  year    = {2009},
  volume  = {106},
  number  = {21},
  pages   = {8417--8422},
  doi     = {10.1073/pnas.0903367106}
}

@article{lachmy2020role,
  title={The role of diabatic heating in {Ferrel cell} dynamics},
  author={Lachmy, Orli and Kaspi, Yohai},
  journal={Geophysical Research Letters},
  volume={47},
  number={23},
  pages={e2020GL090619},
  year={2020},
  publisher={Wiley Online Library}
}

@article{HeldZuritaGotor2025,
  author  = {Held, I. M. and Zurita-Gotor, P.},
  title   = {Misuse of {Kuo--Eliassen} Equation in Studies of the Climatological Mean Meridional Circulation},
  journal = {Journal of the Atmospheric Sciences},
  volume  = {82},
  number  = {8},
  pages   = {1763--1766},
  year    = {2025},
  doi     = {10.1175/JAS-D-24-0246.1}
}

@article{mitas2006recent,
  title={Recent behavior of the {Hadley} cell and tropical thermodynamics in climate models and reanalyses},
  author={Mitas, Christos M and Clement, Amy},
  journal={Geophysical Research Letters},
  volume={33},
  number={1},
  year={2006},
  publisher={Wiley Online Library}
}

@incollection{kalnay2018ncep,
  title={The {NCEP/NCAR} 40-year reanalysis project},
  author={Kalnay, E. and Kanamitsu, M. and Kistler, Robert and Collins, William and Deaven, Dennis and Gandin, Lev and Iredell, Mark and Saha, Suranjana and White, Glenn and Woollen, John and others},
  booktitle={Renewable Energy},
  pages={Vol1\_146--Vol1\_194},
  year={2018},
  publisher={Routledge}
}

@article{solodoch2011excitation,
  title={Excitation of intraseasonal variability in the equatorial atmosphere by {Yanai} wave groups via {WISHE}-induced convection},
  author={Solodoch, Aviv and Boos, William R and Kuang, Zhiming and Tziperman, Eli},
  journal={Journal of the Atmospheric Sciences},
  volume={68},
  number={2},
  pages={210--225},
  year={2011}
}

@article{hersbach2020era5,
  title={The {ERA5} global reanalysis},
  author={Hersbach, Hans and Bell, Bill and Berrisford, Paul and Hirahara, Shoji and Hor{\'a}nyi, Andr{\'a}s and Mu{\~n}oz-Sabater, Joaqu{\'\i}n and Nicolas, Julien and Peubey, Carole and Radu, Raluca and Schepers, Dinand and others},
  journal={Quarterly Journal of the Royal Meteorological Society},
  volume={146},
  number={730},
  pages={1999--2049},
  year={2020},
  publisher={Wiley Online Library}
}

@article{SchwendikeBerryEtAl2021,
  author       = {Schwendike, Juliane and Berry, Gareth J. and Fodor, Katherine and Reeder, Michael J.},
  title        = {On the Relationship Between the {Madden–Julian Oscillation} and the {Hadley and Walker} Circulations},
  journal      = {Journal of Geophysical Research: Atmospheres},
  volume       = {126},
  number       = {4},
  pages        = {e2019JD032117},
  year         = {2021},
  doi          = {10.1029/2019JD032117},
}

@article{wheeler2004all,
  title={An all-season real-time multivariate {MJO} index: Development of an index for monitoring and prediction},
  author={Wheeler, Matthew C and Hendon, Harry H},
  journal={Monthly Weather Review},
  volume={132},
  number={8},
  pages={1917--1932},
  year={2004}
}

@article{LinEtAl2008,
  author  = {Lin, Jialin and Mapes, Brian E. and Han, Weiqing},
  title   = {What Are the Sources of Mechanical Damping in {Matsuno--Gill}-Type Models?},
  journal = {Journal of Climate},
  volume  = {21},
  number  = {2},
  pages   = {165--179},
  year    = {2008},
  doi     = {10.1175/2007JCLI1546.1}
}

@article{yang2011mjo,
  author  = {Yang, Da and Ingersoll, Andrew P.},
  title   = {Testing the hypothesis that the {MJO} is a mixed {Rossby--gravity }wave packet},
  journal = {Journal of the Atmospheric Sciences},
  year    = {2011},
  volume  = {68},
  number  = {2},
  pages   = {226--239},
  doi     = {10.1175/2010JAS3563.1}
}

@article{bretherton1999effective,
  title={The effective number of spatial degrees of freedom of a time-varying field},
  author={Bretherton, Christopher S and Widmann, Martin and Dymnikov, Vladimir P and Wallace, John M and Blad{\'e}, Isabel},
  journal={Journal of Climate},
  volume={12},
  number={7},
  pages={1990--2009},
  year={1999}
}

@article{takasuka2019mrg,
  author  = {Takasuka, Daisuke and Satoh, Masaki and Yokoi, Satoshi},
  title   = {Observational evidence of mixed {Rossby--gravity }waves as a driving force for the {MJO} convective initiation and propagation},
  journal = {Geophysical Research Letters},
  year    = {2019},
  volume  = {46},
  number  = {10},
  pages   = {5546--5555},
  doi     = {10.1029/2018GL081687}
}

@article{takasuka2021mjo,
  author  = {Takasuka, Daisuke and Satoh, Masaki and Yokoi, Satoshi},
  title   = {{MJO} initiation triggered by amplification of upper-tropospheric dry mixed {Rossby--gravity} waves},
  journal = {Geophysical Research Letters},
  year    = {2021},
  volume  = {48},
  number  = {20},
  pages   = {e2021GL094239},
  doi     = {10.1029/2021GL094239}
}

@article{raiter2020tropical,
  title={The tropical atmospheric conveyor belt: A coupled {Eulerian-Lagrangian} analysis of the large-scale tropical circulation},
  author={Raiter, Dana and Galanti, Eli and Kaspi, Yohai},
  journal={Geophysical Research Letters},
  volume={47},
  number={10},
  pages={e2019GL086437},
  year={2020},
  publisher={Wiley Online Library}
}

@article{ventrice2013modified,
  title={A modified multivariate {Madden--Julian Oscillation} index using velocity potential},
  author={Ventrice, Michael J and Wheeler, Matthew C and Hendon, Harry H and Schreck III, Carl J and Thorncroft, Chris D and Kiladis, George N},
  journal={Monthly Weather Review},
  volume={141},
  number={12},
  pages={4197--4210},
  year={2013}
}

@article{lafleur2015some,
  title={Some climatological aspects of the {Madden--Julian} {Oscillation (MJO)}},
  author={Lafleur, Donald M and Barrett, Bradford S and Henderson, Gina R},
  journal={Journal of Climate},
  volume={28},
  number={15},
  pages={6039--6053},
  year={2015}
}

@article{galanti2022spatial,
  title={Spatial patterns of the tropical meridional circulation: Drivers and teleconnections},
  author={Galanti, Eli and Raiter, Dana and Kaspi, Yohai and Tziperman, Eli},
  journal={Journal of Geophysical Research: Atmospheres},
  volume={127},
  number={2},
  pages={e2021JD035531},
  year={2022},
  publisher={Wiley Online Library}
}

@article{Arnold2012,
  author  = {Arnold, N. P. and Tziperman, E. and Farrell, B. F.},
  title   = {Abrupt transition to strong superrotation driven by equatorial wave resonance in an idealized {GCM}},
  journal = {Journal of the Atmospheric Sciences},
  year    = {2012},
  volume  = {69},
  pages   = {626--640},
  doi     = {10.1175/JAS-D-11-0136.1}
}

@article{kiladis2014comparison,
  title={A comparison of {OLR} and circulation-based indices for tracking the MJO},
  author={Kiladis, George N and Dias, Juliana and Straub, Katherine H and Wheeler, Matthew C and Tulich, Stefan N and Kikuchi, Kazuyoshi and Weickmann, Klaus M and Ventrice, Michael J},
  journal={Monthly Weather Review},
  volume={142},
  number={5},
  pages={1697--1715},
  year={2014}
}

@article{raiter2024linking,
  title={Linking future tropical precipitation changes to zonally-asymmetric large-scale meridional circulation},
  author={Raiter, Dana and Galanti, Eli and Chemke, Rei and Kaspi, Yohai},
  journal={Geophysical Research Letters},
  volume={51},
  number={6},
  pages={e2023GL106072},
  year={2024},
  publisher={Wiley Online Library}
}

@article{li2025distinct,
  title={Distinct intraseasonal oscillation modes over the tropical {Indo-Pacific Oceans}},
  author={Li, Yiran and Hu, Haibo and Liu, Fei and Patterson, Matthew and Yang, Xiu-Qun and Lu, Kecheng and Mao, Kefeng and Wang, Ziyi and Wang, Rongrong},
  journal={Geophysical Research Letters},
  volume={52},
  number={9},
  pages={e2024GL113263},
  year={2025},
  publisher={Wiley Online Library}
}

@article{takasuka2021diversity,
  title={Diversity of the {Madden--Julian} oscillation: Initiation region modulated by the interaction between the intraseasonal and interannual variabilities},
  author={Takasuka, Daisuke and Satoh, Masaki},
  journal={Journal of Climate},
  volume={34},
  number={6},
  pages={2297--2318},
  year={2021}
}

@article{dasgupta2020exploring,
  title={Exploring the long-term changes in the {Madden Julian Oscillation} using machine learning},
  author={Dasgupta, Panini and Metya, Abirlal and Naidu, CV and Singh, Manmeet and Roxy, MK},
  journal={Scientific Reports},
  volume={10},
  number={1},
  pages={18567},
  year={2020},
  publisher={Nature Publishing Group UK London}
}

@book{krishnamurti1979tropical,
  title={Tropical Meteorology},
  author={Krishnamurti, Tiruvalam N. and Stefanova, Lydia and Misra, Vasubandhu},
  year={1979},
  publisher={Springer}
}

@article{Carlson2016,
  author  = {Carlson, H. and Caballero, R.},
  title   = {Enhanced {MJO} and transition to superrotation in warm climates},
  journal = {Journal of Advances in Modeling Earth Systems},
  year    = {2016},
  volume  = {8},
  pages   = {304--318},
  doi     = {10.1002/2015MS000615}
}

@article{SobelMaloney2012,
  author  = {Sobel, Adam H. and Maloney, Eric D.},
  title   = {An Idealized Semi-Empirical Framework for Modeling the {Madden--Julian Oscillation}},
  journal = {Journal of the Atmospheric Sciences},
  year    = {2012},
  volume  = {69},
  number  = {5},
  pages   = {1691--1705},
  doi     = {10.1175/JAS-D-11-0118.1}
}

@article{SobelMaloney2013,
  author  = {Sobel, Adam H. and Maloney, Eric D.},
  title   = {Moisture Modes and the Eastward Propagation of the {MJO}},
  journal = {Journal of the Atmospheric Sciences},
  year    = {2013},
  volume  = {70},
  number  = {1},
  pages   = {187--192},
  doi     = {10.1175/JAS-D-12-0189.1}
}

@article{AdamesKim2016,
  author  = {Adames, {\'A}ngel F. and Kim, Daehyun},
  title   = {The {MJO} as a Dispersive, Convectively Coupled Moisture Wave: Theory and Observations},
  journal = {Journal of the Atmospheric Sciences},
  year    = {2016},
  volume  = {73},
  number  = {3},
  pages   = {913--941}
}

@article{YangIngersoll2013,
  author  = {Yang, Da and Ingersoll, Andrew P.},
  title   = {Triggered Convection, Gravity Waves, and the {MJO}: A Shallow-Water Model},
  journal = {Journal of the Atmospheric Sciences},
  year    = {2013},
  volume  = {70},
  number  = {8},
  pages   = {2476--2486},
  doi     = {10.1175/JAS-D-12-0255.1}
}

@article{SchneiderEtAl2014,
  author  = {Schneider, Tapio and Bischoff, Tobias and Haug, Gerald H.},
  title   = {Migrations and dynamics of the intertropical convergence zone},
  journal = {Nature},
  volume  = {513},
  number  = {7516},
  pages   = {45--53},
  year    = {2014},
  doi     = {10.1038/nature13636}
}

@article{YaoLuLiu2026,
  author  = {Yao, Wuqiushi and Lu, Jianhua and Liu, Yimin},
  title   = {Rotational flow dominates abrupt seasonal change in zonally asymmetric tropical meridional circulation},
  journal = {Geophysical Research Letters},
  year    = {2026},
  volume  = {53},
  pages   = {e2025GL118924},
  doi     = {10.1029/2025GL118924}
}

@article{ZaplotnikEtAl2022,
  author  = {Zaplotnik, {\v{Z}}iga and Pikovnik, Matic and Boljka, Luka},
  title   = {Recent{ Hadley} circulation strengthening: a trend or multidecadal variability?},
  journal = {Journal of Climate},
  volume  = {35},
  number  = {13},
  pages   = {4157--4176},
  year    = {2022},
  doi     = {10.1175/JCLI-D-21-0204.1}
}

@article{YingEtAl2024,
  author  = {Ying, Tian and Li, Jianping and Fu, Qiang and Liu, Guoshun and Zhang, Lei and Xia, Yan and Hu, Yongyun},
  title   = {Fractional change of scattering and absorbing aerosols contributes to Northern Hemisphere Hadley circulation expansion},
  journal = {Science Advances},
  volume  = {10},
  number  = {46},
  pages   = {eadq9716},
  year    = {2024},
  doi     = {10.1126/sciadv.adq9716}
}

@article{su2025consistency,
  title={Consistency of Changes in the Ascending and Descending Positions of the {Hadley Circulation} Using Different Methods},
  author={Su, Qianye and Liu, Chunlei and Zhang, Yu and Qiu, Juliao and Li, Jiandong and Xue, Yufeng and Cao, Ning and Liao, Xiaoqing and Yang, Ke and Zheng, Rong and others},
  journal={Atmosphere},
  volume={16},
  number={4},
  pages={367},
  year={2025},
  publisher={MDPI}
}

\newpage
\newpage
\newpage

\section*{Acknowledgments}

We thank Eli Tziperman and Huayu Chen for insightful discussions. We also thank Qianye Su for help with diagnosing the Kuo–Eliassen equation, and Lior Hochman for downloading the ERA5 tendency datasets.

%
%


%
%
%
%
%

\end{document}